\documentclass[conference]{IEEEtran}
\normalsize

\ifCLASSINFOpdf

\else

\fi
\usepackage{cite}
\usepackage{graphicx}
\usepackage{psfrag}
\usepackage{subfigure}
\usepackage{url}
\usepackage{stfloats}
\usepackage{amsmath}
\usepackage{array}
\usepackage{amssymb}
\usepackage{makecell}
\usepackage{stfloats}
\usepackage{booktabs}
\usepackage{threeparttable}
\usepackage{multirow}
\usepackage{float}

\begin{document}

\title{Capacity Bounds for Bandlimited Gaussian Channels With Peak-to-Average-Power-Ratio Constraint}

\author{\IEEEauthorblockN{Yizhu~Wang, Jing~Zhou, and~Wenyi~Zhang}
\IEEEauthorblockA{CAS Key Laboratory of Wireless-Optical Communications\\University of Science and Technology of China, Hefei, China\\
Email: wyz94@mail.ustc.edu.cn; jzee@ustc.edu.cn; wenyizha@ustc.edu.cn}}

\maketitle

\begin{abstract}
We revisit Shannon's problem of bounding the capacity of bandlimited Gaussian channel (BLGC) with peak power constraint,
and extend the problem to the peak-to-average-power-ratio (PAPR) constrained case.
By lower bounding the achievable information rate of pulse amplitude modulation with independent and identically distributed input under a PAPR constraint,
we obtain a general capacity lower bound with respect to the shaping pulse.
We then evaluate and optimize the lower bound by employing some parametric pulses, thereby improving the best existing result.
Following Shannon's approach, capacity upper bound for PAPR constrained BLGC is also obtained.
By combining our upper and lower bounds, the capacity of PAPR constrained BLGC is bounded to within a finite gap which tends to zero as the PAPR constraint tends to infinity.
Using the same approach, we also improve existing capacity lower bounds for bandlimited optical intensity channel at high SNR.
\end{abstract}
\IEEEpeerreviewmaketitle

\section{Introduction}
In Part IV of his 1948 landmark paper \cite{Shannon48}, Shannon introduced a bandlimited Gaussian channel (BLGC) as
\begin{equation}
\label{BLGC}
Y(t)=X(t)+Z(t),
\end{equation}
where $X(t)$ is bandlimited to $\mathcal{W}\triangleq[-W,W]$ Hz,
and $Z(t)$ is additive white Gaussian noise (AWGN) of power spectral density $N_0$ on $\mathcal{W}$.
Two types of input constraint were considered:
1) an average power (AP) constraint $P$, and 2) a peak power (PP) or amplitude constraint as
$|X(t)|\leq A$.
For brevity we denote these two cases as AP-BLGC and PP-BLGC, respectively,
and use similar notations for other kinds of channels subsequently.
For an AP-BLGC of signal-to-noise-ratio $\mathrm{SNR}\triangleq\frac{P}{N_0 W}$, Shannon established the famous formula
\begin{equation}
\label{Wlog}
C=W\log_2\left(1+\mathrm{SNR}\right)\mspace{4mu} \textrm{bits/second}.
\end{equation}
But for PP-BLGC, Shannon only provided upper and lower bounds on its capacity, and almost no further results have been known since then. An exception is \cite{Shamai88},
which improved Shannon's capacity lower bound for PP-BLGC by 0.567 dB.

A related topic is the capacity problem of the PP constrained discrete-time Gaussian channel (DTGC) as
\begin{equation}
\label{DTGC}
Y_n=X_n+Z_n,
\end{equation}
where the channel input satisfies $|X_n|\leq A$ (possibly combined with an AP constraint as $\lim\limits_{N\to\infty}\frac{1}{N}\mathbf E\left[\sum_{n=1}^{N}x_n^2\right]\leq P$), 
and the Gaussian noise $Z_n\sim \mathcal{N}(0,\sigma_z^{2})$ is memoryless.
The capacity of PP-DTGC can be numerically evaluated using techniques initiated in \cite{Smith71}.
However, PP-DTGC only models intersymbol-interference-free (ISI-free) transmission, which cannot achieve optimal bandwidth efficiency in BLGC with PP constraint.
The reason is that, for a Nyquist shaping pulse, a smaller excess bandwidth causes larger sidelobes,
whose superposition sharply increases the peak of continuous-time signal waveform as the excess bandwidth tends to zero \cite{BIM}.

In this paper, we revisit the continuous-time scenario and consider an extended version of PP-BLGC, namely,
BLGC with a peak-to-average-power-ratio (PAPR) constraint.
In Sec. II we establish a general capacity lower bound for PAPR-BLGC by lower bounding the information rate of constrained pulse amplitude modulation (PAM) with independent and identically distributed (i.i.d.). input symbols.
The bound has a form similar to (\ref{Wlog})
except for a pre-SNR factor determined by the shaping pulse.
We maximize the pre-SNR factor by optimizing the pulse, and the tightest lower bound obtained outperforms the result on PP-BLGC in \cite{Shamai88} by 0.926 dB.
Moreover, our bound reduces to (\ref{Wlog}) when the PAPR constraint tends to infinity.
Capacity upper bound for PAPR-BLGC is also provided based on Shannon's approach for PP-BLGC in \cite{Shannon48}.
In Sec. III, using the same techniques, we study bandlimited optical intensity channels (BLOIC) \cite{BIM,HK04,ZZ17},
improve capacity lower bounds in \cite{ZZ17} at high SNR, and disprove a conjecture therein.
Our result also outperforms the capacity lower bound of AP-BLOIC in \cite{HJC16} obtained by a sphere packing argument.
Finally, Sec. IV concludes the paper.

\section{Capacity Bounds for PAPR-BLGC}
Following \cite{Shannon48}, the capacity of a BLGC is defined as
\begin{equation}
\label{eqn:CBC1}
 C=2W\cdot\mathop {\sup }\limits_{p_{\bf X}({\bf x})} {\cal I}[X(t);Y(t)]
\end{equation}
where
\begin{flalign}
\label{eqn:CBC}
{\cal I}&[X(t);Y(t)]\notag\\
=&\mathop {\lim}\limits_{N\to \infty } \frac{1}{2N+1}\int\!\!\!\int {p_{\bf {X,Y}}({\bf x},{\bf y})\log \frac{p_{\bf {X,Y}}({\bf x},{\bf y})}{p_{\bf {X}}({\bf x})p_{\bf {Y}}({\bf y})}} \mathrm{d}{\bf x}\mathrm{d}{\bf y}
\end{flalign}
is the mutual information per degree-of-freedom (DoF) between $X(t)$ and $Y(t)$,
${\rm {\bf X}}=\left[X[-N], \ldots, X[0], \ldots, X[N] \right]$ and ${\rm {\bf Y}}=\left[Y[-N],\ldots,Y[0], \ldots,  Y[N]\right]$
are Nyquist sample sequences of $X(t)$ and $Y(t)$, respectively.
For a PAPR-BLGC, the constraint on the input waveform $X(t)$ is
\begin{equation}
\label{P}
\lim_{T\rightarrow\infty}\frac{1}{2T} \mathbf E\left[\int_{-T}^T {X^2(t)\mathrm{d}t}\right] \leq P, \mspace{8mu} X^2(t)\leq rP,
\end{equation}
where $r>0$. We refer to (\ref{P}) as a PAPR constraint $r$.

In \cite{Shannon48}, by deriving high and low peak-to-noise ratio (PNR) asymptotic results, Shannon showed that
the capacity of PP-BLGC with $\mathrm{PNR}=\frac{A^2}{N_0W}$ can be expressed as
\begin{equation}
\label{PPC}
C_\textrm{PP-BLGC}=W\log_2\left(1+\eta(\mathrm{PNR})\cdot\mathrm{PNR}\right)\mspace{4mu} \textrm{bits/second},
\end{equation}
where $\eta(\mathrm{PNR})$ is an unknown pre-PNR factor satisfying $\lim_{\mathrm{PNR}\to 0}\eta(\mathrm{PNR})=1$. When $r$ is sufficiently small, the PP constraint dominates and the PAPR-BLGC behaves like a PP-BLGC. When $r$ tends to infinity the PAPR-BLGC behaves like an AP-BLGC. So we can infer that the capacity of PAPR-BLGC can be written as
\begin{equation}
C_\textrm{PAPR-BLGC}=W\log_2\left(1+\eta(\mathrm{SNR},r)\cdot\mathrm{SNR}\right)\mspace{4mu} \textrm{bits/second}.
\end{equation}
The pre-SNR factor $\eta(\mathrm{SNR},r)$ is a non-decreasing function of $r$ satisfying $\lim\limits_{r\to\infty}\eta(\mathrm{SNR},r)=1$ and $\lim\limits_{r\to 0}\eta(\mathrm{SNR},r)=0$.

\subsection{Capacity Lower Bound}
We derive capacity lower bounds for PAPR-BLGC by lower bounding the information rate of PAM waveform ensemble like
\begin{equation}
\label{XtXn}
X(t)=\sum \limits _{n} {X_n g\left(t-nT_\textrm{s}\right)}
\end{equation}
 under the PAPR constraint, where the shaping pulse $g(t)$ is an ${\cal {L}}_2$ (i.e., finite-energy) funtion normalized as
\begin{equation}
\label{g}
E_g=\int _{-\infty} ^{\infty} g^2(t)\mathrm d t =\frac{1}{2W}.
\end{equation}
We denote the Fourier transform of $g(t)$ by $\hat{g}(f)$.
Let the input symbols $\{X_n\}$ be i.i.d. satisfying
$\mathbf E[X_n]=0$, $\mathbf E[|X_n|^2]=\sigma_X^2$.
To achieve the maximum pre-log factor, we let $g(t)$ be bandlimited to $\cal W$ and $T_\textrm{s}=\frac{1}{2W}$.
Since the power of an i.i.d. PAM ensemble as (\ref{XtXn}) is given by $\frac{1}{T_\textrm{s}}E_g\sigma_X^2$ [\ref{Lapidoth}, Sec. 14.5.1],
to meet the AP constraint in (\ref{P}) we let $\sigma_X^2\leq P$.
To meet the PP constraint in (\ref{P}), first we let $X_n$ satisfy $|X_n|\leq a$. Define
\begin{equation}
\label{S}
{\cal S}\triangleq\max\limits_{t\in\big[0,\frac{1}{2W}\big)}\sum\limits_{i=-\infty}^{\infty}\left|g\left(t-\frac{i}{2W}\right)\right|,
\end{equation}
which evaluates the maximum possible superposition of peak caused by pulse shaping.
Then $|X(t)|\leq a{\cal S}$ and the PP constraint in (\ref{P}) is equivalent to $a=\frac{\sqrt{rP}}{{\cal S}}$.

We use the following lemma to derive our lower bound. The lemma is essentially due to Shannon \cite{Shannon48}, and
can also be proved using Szeg\"{o}'s theorem \cite{Shamai88}.

\emph{Lemma 1}:
The achievable information rate of the i.i.d. PAM ensemble (\ref{XtXn}) is lower bounded by
\begin{equation}
\label{Lemma}
I_{\textrm{PAM}} \ge W\log_2 \left(1+\frac{\mathcal G \exp{(2h(X))}}{2 \pi e N_0 W}\right)\mspace{4mu} \textrm{bits/second}
\end{equation}
where $h(X)$ is the differential entropy of $X_n$, and
\begin{equation}
\label{G}
\mathcal G\triangleq\exp\left(\frac{1}{W}\int_{0}^{W} {\log\left|2W\cdot {\hat{g}(f)} \right|^2 \mathrm{d}f}\right).
\end{equation}

\begin{table*}[tbp]
\renewcommand\arraystretch{1}
\centering
\caption{List of Pulses Used ($\beta \in (0,1]$ for all cases)}
\scalebox{1}
{
\begin{tabular}{l|l|l}
\hline
\makecell[cc]{\textbf{Name}}&\makecell[cc]{\textbf{Definition}}&\makecell[cc]{\textbf{Remark}}\\ \hline
S2 &\makecell[lc]{$g(t)=\frac{\sqrt{3}}{2}\mathrm{sinc}^2(Wt)$, $\mathrm{sinc}(t)\triangleq\frac{\sin \pi t}{\pi t}$}&\makecell[lc]{Square of sinc pulse, used in \cite{Shannon48}.} \\\hline
\makecell[lc]{SC} &\makecell[lc]{$g(t)=\sqrt{2}\left(\mathrm {sinc}\left(2Wt-\frac{1}{2}\right)+\mathrm {sinc}\left(2Wt+\frac{1}{2}\right)\right)$}& \makecell[lc]{Spectral-cosine pulse, used in \cite{Shamai88},\\also called duobinary pulse.}\\\hline
RC &$g(t)=\frac{2}{\sqrt{-\beta^2+3\beta+4}}\mathrm {sinc}\left(\frac{2W t}{1+\beta}\right)\frac{\cos\left(\frac{2\pi \beta Wt}{1+\beta}\right)}{1-\left(\frac{4\beta Wt}{1+\beta}\right)^2} $&\makecell[lc]{Raised cosine filter \cite{Lapidoth}.}\\\hline
PL &\makecell[lc]{$g(t)=\sqrt{\frac{3}{-\beta^2+2\beta+3}}\mathrm{sinc}\left(\frac{2Wt}{1+\beta}\right)\mathrm{sinc}\left(\frac{2\beta Wt}{1+\beta}\right)$}& \makecell[lc]{Parametric linear pulse \cite{PL}.}\\\hline
BTN &\makecell[lc]{$g(t)=\sqrt{\frac{1}{-0.36\beta^2+0.64\beta +1}}\mathrm{sinc}\left(\frac{2Wt}{1+\beta}\right)\frac{\frac{4W\pi\beta t}{\ln2(1+\beta)}\sin\left(\frac{2W\pi\beta t}{1+\beta}\right)+2\cos \frac{2W\pi \beta t}{1+\beta}-1}{\left(\frac{2W\pi\beta t}{\ln2 (1+\beta)}\right)^2+1}$} &``Better than Nyquist'' pulse \cite{BTN}.\\\hline
ICIT&\makecell[lc]{$\hat{g}(f)=
\begin{cases}
\frac{1}{2W},\mspace{412mu}|f|\leq\frac{1-\beta}{1+\beta}W\\
\frac{1}{2W}\left(1-\frac{1}{2\gamma}\mathrm{arccos}\left(\mathrm{arctan}\left(\mathrm{tan}(1)\frac{1+\beta}{2\beta W}\left(W-|f|\right)\right)\right)\right),\mspace{20mu}\frac{1-\beta}{1+\beta}W<|f|\leq\frac{W}{1+\beta}\\
\frac{1}{4\gamma W}\mathrm{arccos}\left(\mathrm{arctan}\left(\mathrm{tan}(1)\frac{1+\beta}{2\beta W}\left(|f|-\frac{1-\beta}{1+\beta}W\right)\right)\right),\mspace{50mu}\frac{W}{1+\beta}<|f|\leq W  \\
0,\mspace{432mu}|f|>W.
\end{cases}$}&\makecell[lc]{Inverse-cosine inverse-tangent \\pulse \cite{ICIT}. No explicit time-\\domain expression is known.}\\\hline
\end{tabular}}
\end{table*}

We tighten (\ref{Lemma}) by finding a maxentropic distribution for $X_n$ under the constraints $\sigma_X^2\leq P$ and $|X_n|\leq\frac{\sqrt{rP}}{{\cal S}}$.
According to [\ref{CT06}, Theorem 12.1.1] we obtain the following solution.

1) When $\frac{r}{{\cal S}^2}> 3$, the maxentropic distribution is a truncated Gaussian distribution with probability density function (PDF)
\begin{equation}
\label{TG}
p_X(x)=\frac{1}{\mathrm{erf}(\lambda)\sqrt{2\pi\sigma^2}}\exp\left(-\frac{x^2}{2\sigma^2}\right),
\mspace{8mu} |x|\leq\sqrt{2}\lambda\sigma,
\end{equation}
where $\mathrm{erf}(\lambda)\triangleq\frac{1}{\sqrt{\pi}}\int_{-\lambda}^{\lambda}e^{-t^2}\mathrm{d}t$,
\begin{equation}
\label{sigma}
\sigma^2=\frac{r}{2\lambda^2 {\cal S}^2}P,
\end{equation}
and the parameter $\lambda>0$ is the unique solution of
\begin{equation}
\label{lambda}
\frac{P}{\sigma^2}=1-\frac{2\lambda}{\sqrt{\pi}\mathrm{erf}(\lambda)\exp(\lambda^2)}.
\end{equation}
The corresponding differential entropy is
\begin{equation}
\label{hTG}
h(X)=\frac{1}{2}\log\left(2\pi e \sigma^2\cdot\mathrm{erf}^2(\lambda)\exp\left(\frac{P}{\sigma^2}-1\right)\right).
\end{equation}

2) When $0<\frac{r}{{\cal S}^2}\leq 3$, the maxentropic distribution is a uniform distribution with PDF
\begin{equation}
p_X(x)=\frac{{\cal S}}{2\sqrt{rP}}, \mspace{8mu} |x|\leq\frac{\sqrt{rP}}{{\cal S}},
\end{equation}
and the corresponding differential entropy is
\begin{equation}
\label{U}
h(X)=\log\frac{2\sqrt{rP}}{{\cal S}}.
\end{equation}
Substituting (\ref{hTG}) and (\ref{U}) into (\ref{Lemma}) yields the following result.

\emph{Proposition 1}: The capacity of the BLGC as (\ref{BLGC}) with a PAPR constraint $r$ given by (\ref{P}), is lower bounded by
\begin{equation}
\label{LB}
C_{\textrm{PAPR-BLGC}}(r)\ge
W\log_2\left(1+\eta_g(r)\mathrm{SNR}\right)\mspace{4mu} \textrm{bits/second},
\end{equation}
where
\begin{equation}
\label{LB1}
\begin{split}
\eta_g(r)=
\begin{cases}
\frac{{\cal G}r}{2e\lambda^2{\cal S}^2} \mathrm{erf}^2(\lambda)\exp\left(\frac{2\lambda^2{\cal S}^2}{r}\right),\mspace{32mu}\frac{r}{{\cal S}^2}>3\\
\frac{ 2{\cal G} r}{\pi e {\cal S}^2},\mspace{170mu}0<\frac{r}{{\cal S}^2}\leq3,
\end{cases}
\end{split}
\end{equation}
${\cal S}$ and $\cal G$ are defined by (\ref{S}) and (\ref{G}), respectively, with respect to a pulse $g(t)$ normalized as (\ref{g}), and $\lambda>0$ is the unique solution of (\ref{lambda}).

\subsection{Pulse Optimization}

\begin{figure}[t]
\centering
\includegraphics[width=3.2in,height=2.4in]{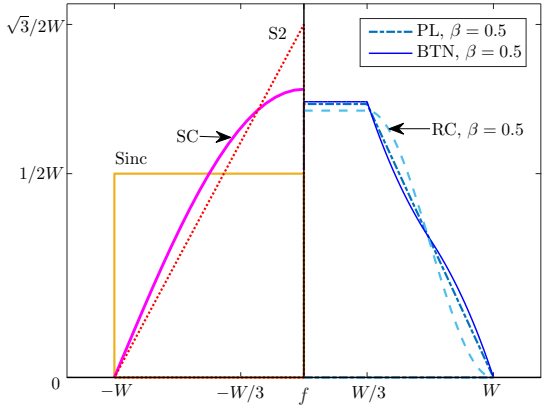}
\caption{Spectra of pulses considered in PAPR-BLGC.\protect\footnotemark}
\label{Pulse}
\end{figure}

The general lower bound (\ref{LB}) can be evaluated using specific pulses.
Table I lists the pulses considered in this paper. Fig. \ref{Pulse} shows the
Fourier transform $\hat{g}(f)$ of the pulses used in PAPR-BLGC, all of which are normalized to satisfy (\ref{g}).
Since no analytic solution for (\ref{lambda}) is known, for our result, the pre-SNR factor (\ref{LB1}) is evaluated numerically.
Although the exact value of ${\cal S}$ is difficult to find analytically for arbitrary pulses,
the sidelobes of the pulses in Table I decay asymptotically as $\frac{1}{t^2}$ or faster as $t\to\infty$. This makes reliable numerical evaluation of ${\cal S}$ possible.
For parametric pulses such as the RC pulse, we optimize it over $\beta\in (0,1]$ to maximize $\eta_g(r)$ for each $r$.

\begin{figure}[t]
\centering
\includegraphics[width=3.2in,height=2.4in]{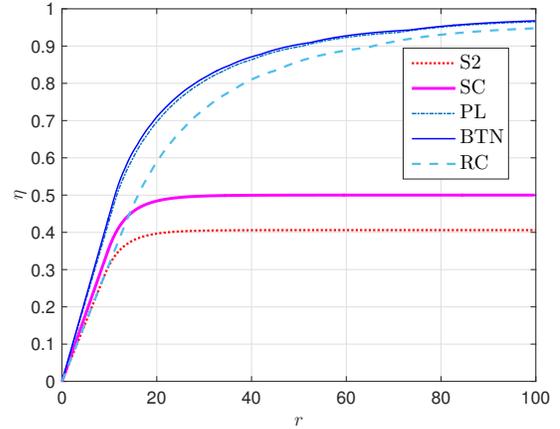}
\caption{Pre-SNR factors of capacity lower bounds of PAPR-BLGC.}
\label{1}
\end{figure}

\begin{figure*}
\begin{minipage}[t]{0.33\linewidth}
\centering
\includegraphics[width=2.48in,height=1.86in]{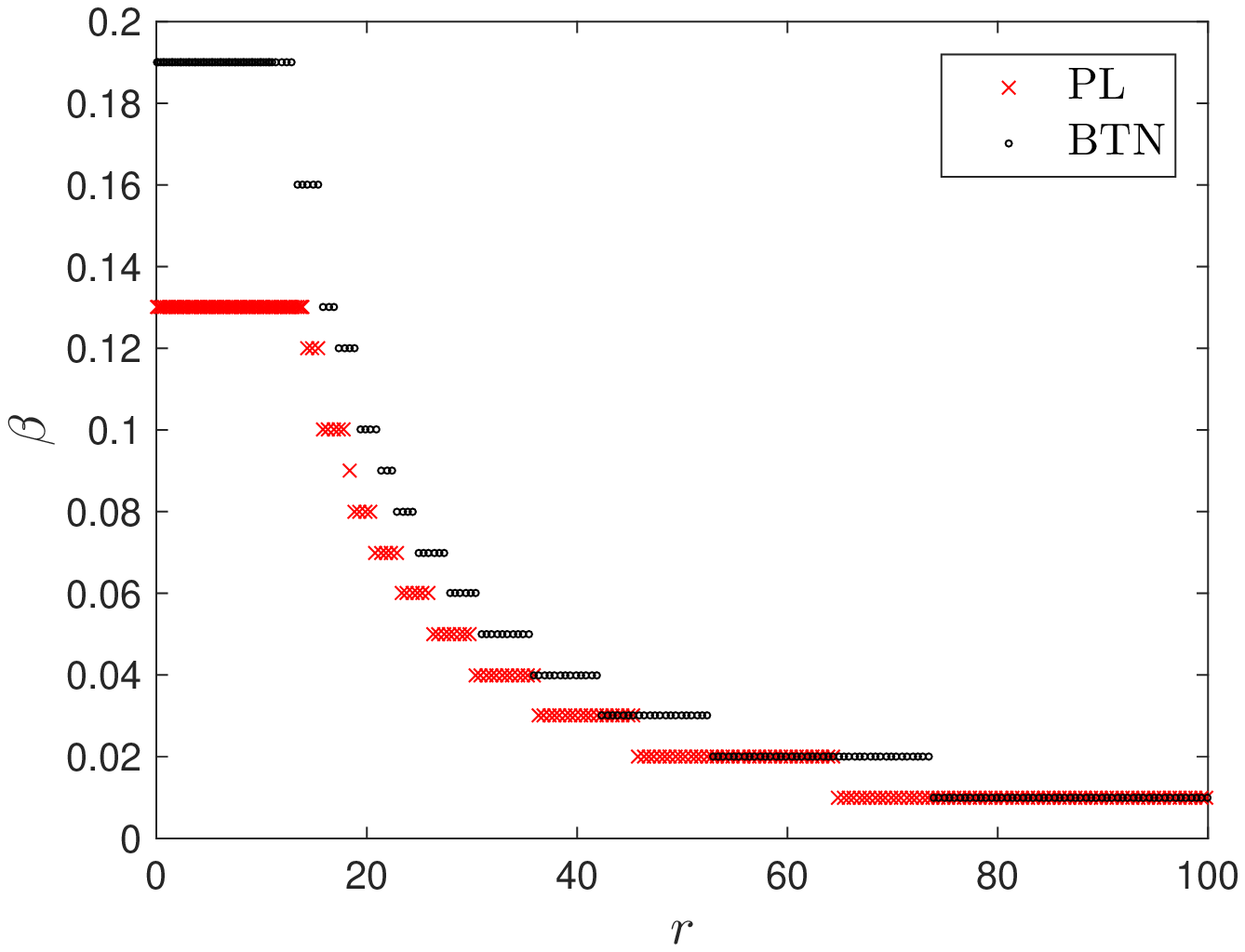}
\caption{Optimal value of $\beta$ for BL and BTN $\mspace{38mu}$ pulses in PAPR-BLGC (the step size for $\mspace{38mu}$ optimizing $\beta$ is 0.01).}
\label{beta}
\end{minipage}
\begin{minipage}[t]{0.33\linewidth}
\centering
\includegraphics[width=2.48in,height=1.86in]{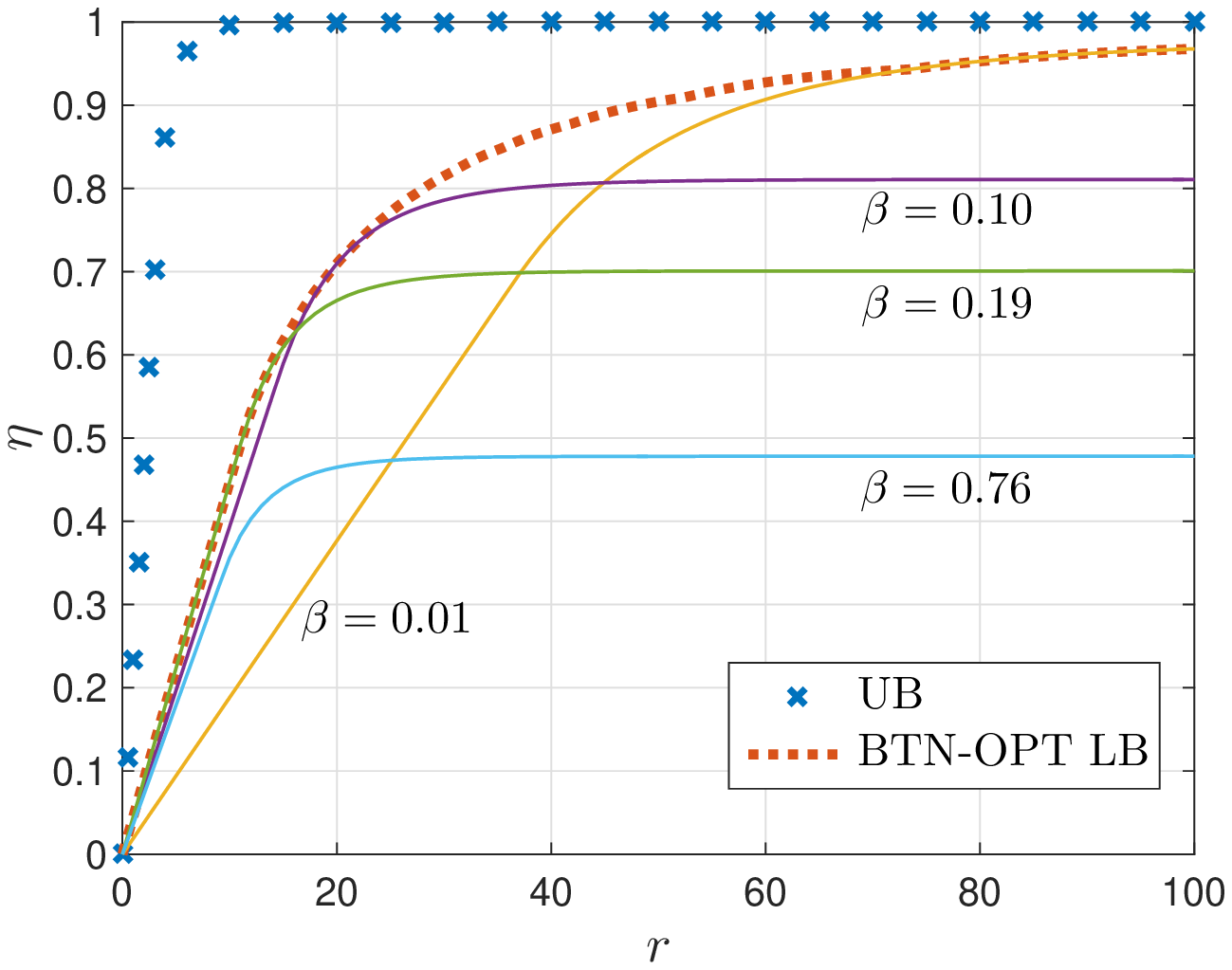}
\caption{Pre-SNR factors for specific BTN pulses.$\mspace{25mu}$}
\label{BTN}
\end{minipage}
\begin{minipage}[t]{0.33\linewidth}
\centering
\includegraphics[width=2.48in,height=1.86in]{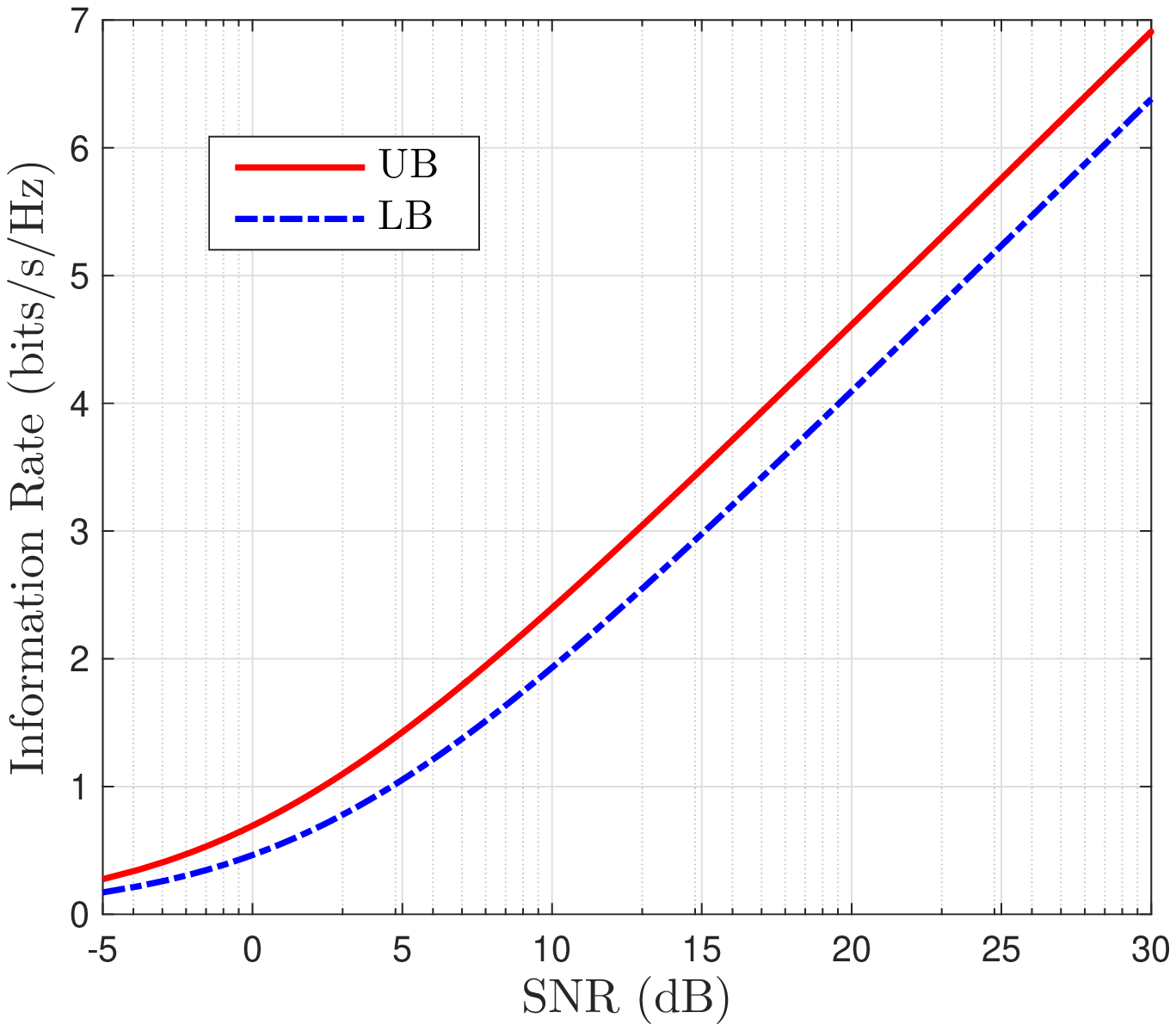}
\caption{Capacity bounds for PAPR-BLGC when $r=14$.}
\label{UBLB}
\end{minipage}
\end{figure*}

In Fig. \ref{1}, our numerical results for the pre-SNR factor lower bound (\ref{LB1}) is provided. The BTN pulse shows the best performance.
Fig. \ref{beta} provides the optimal value of $\beta$ for the BTN pulse and the PL pulse for each $r$ (the step size for optimizing $\beta$ is 0.01).
Fig. \ref{BTN} provides the pre-SNR factor lower bounds obtained by using the BTN pulses with some specific values of $\beta$, and
the best lower bound obtained by optimizing the BTN pulse over $\beta\in (0,1]$ (denoted ``BTN-OPT LB'').
The optimal value $\beta^*$ decreases as the PAPR constraint $r$ increases.
As $r\to\infty$, it appears that $\beta^*$ tends to zero for both the BTN pulse and the PL pulse, i.e. both pulses tend to the sinc pulse.
The following result shows that the lower bounds for the BTN, PL and RC pulses in Fig. \ref{1} tend to one as $r\to\infty$.

\footnotetext{For clarity, we only show a half of the spectrum of each pulse here (and in Fig. \ref{PulseOIC}).
Note that the spectrum of them are symmetric with respect to $f=0$.}

\emph{Proposition 2}: For the PAPR-BLGC, as $r$ grows without bound, there exists a parametric pulse $g(t)$ such that
 $\eta_g(r)$ tends to one and (\ref{LB}) tends to (\ref{Wlog}).

\emph{Outline of Proof}:
By substituting (\ref{sigma}) into (\ref{lambda}) we obtain
\begin{equation}
\label{RHS}
\frac{r}{2{\cal S}^2}=\lambda^2+\frac{2\lambda^3}{\sqrt{\pi}\mathrm{erf}(\lambda)\exp(\lambda^2)-2\lambda}.
\end{equation}
By noting that the RHS of (\ref{RHS}) is continuous for $0<\lambda<\infty$ and $\lim\limits_{\lambda\to 0}\frac{\mathrm{erf}(\lambda)}{\lambda}=\frac{2}{\sqrt{\pi}}$, we can prove that when ${\cal S}$ is finite, if $r$ grows without bound, then $\lambda$ will also grow without bound, and $\frac{2\lambda^2{\cal S}^2}{r}\to 1$.
Thus when $\frac{r}{{\cal S}^2}>3$, we have $\lim\limits_{r\to\infty}\eta_g(r)={\cal G}$ for a given $g(t)$.
Using a parametric pulse which tends to the sinc pulse as $\beta\to 0$ (e.g., the BTN pulse),
we can make $\eta_g(r)$ arbitrarily close to one as $r$ grows without bound.
\hfill{\small$\blacksquare$}

By replacing $rP$ in the second case of Proposition 1 with $A^2$, we revisit a capacity lower bound for PP-BLGC established in \cite{Shamai88}, which is equivalent to
\begin{equation}
\label{ShamaiLB}
\eta(\mathrm{PNR})\ge\frac{2{\cal G}}{\pi e {\cal S}^2},
\end{equation}
where $\eta(\mathrm{PNR})$ is the pre-PNR factor in (\ref{PPC}). Thus we can compare our results with known capacity lower bounds for PP-BLGC.
Using the S2 pulse, in \cite{Shannon48} it was shown that
\begin{equation}
\label{SLB}
\eta(\mathrm{PNR})\ge\frac{2}{\pi e^3}\approx0.03170;
\end{equation}
Using the SC pulse, in \cite{Shamai88}
\begin{equation}
\label{SmLB}
\eta(\mathrm{PNR})\ge\frac{\pi}{32e}\approx0.03612
\end{equation}
was obtained, which improved (\ref{SLB}) by 0.567 dB.
According to Fig. 2, both the BTN and PL pulses outperform the SC pulse.
In particular, employing the BTN pulse we obtain
\begin{equation}
\label{nLB}
\eta(\mathrm{PNR})\ge0.04470.
\end{equation}
This tightens the existing PP-BLGC result (\ref{SmLB}) by $10\log_{10}\frac{0.04470}{0.03612}\approx0.926$ dB.

\emph{Remark}: In PP-BLGC, the achievable rate using a specific pulse is determined by $\frac{{\cal G}}{{\cal S}^2}$, which is proportional to the slope of the corresponding curve in Fig. \ref{BTN} when $r\leq 3{\cal S}^2$.
But for PAPR-BLGC, when $r> 3{\cal S}^2$, a pulse with larger $\frac{{\cal G}}{{\cal S}^2}$ is not necessarily better. Nevertheless, if two pulses have the same $\cal G$ ($\cal S$), we always prefer the one that with smaller $\cal S$ (larger $\cal G$).

\subsection{Capacity Upper Bound}
In \cite{Shannon48}, by proving an asymptotic upper bound for the PP-BLGC as
$\lim_{\mathrm{PNR}\to\infty}\eta(\mathrm{PNR})\leq\frac{2}{\pi e}$,
Shannon essentially established the following general upper bound for the capacity of bandlimited channels with an amplitude constraint.

\emph{Lemma 2}: Let the input of a BLGC be constrained by
an amplitude constraint $\underline{A}\leq X(t)\leq \overline{A}$, where $-\infty\leq\underline{A}<\overline{A}\leq\infty$. The capacity of this channel is upper bounded by
\begin{equation}
\label{DT2BL}
C_{\textrm {BLGC}}\leq {C}_{\textrm {DTGC}}\cdot 2W \mspace{10mu} \textrm {bits/second},
\end{equation}
where $C_{\textrm {DTGC}}$ is the capacity of a DTGC as (\ref{DTGC}) satisfying $\underline{A}\leq X_n\leq \overline{A}$ and $Z_n\sim\mathcal N(0,N_0W)$.

By combining this and the numerical computation technique in \cite{Smith71}, we provide numerical upper bound (denoted ``UB'') for the PAPR-BLGC in Fig. \ref{BTN}. It is shown that the gap between our capacity upper and lower bounds
is still large for small $r$, and the gap decreases as $r$ increases. For example, when $r=14$, the high SNR gap is about 2.30 dB; see Fig. 5.
As $r\to\infty$ the gap diminishes, as indicated by Proposition 2.

\section{Capacity Bounds for PAPR-BLOIC}
\begin{figure*}
\begin{minipage}[t]{0.33\linewidth}
\centering
\includegraphics[width=2.48in,height=1.86in]{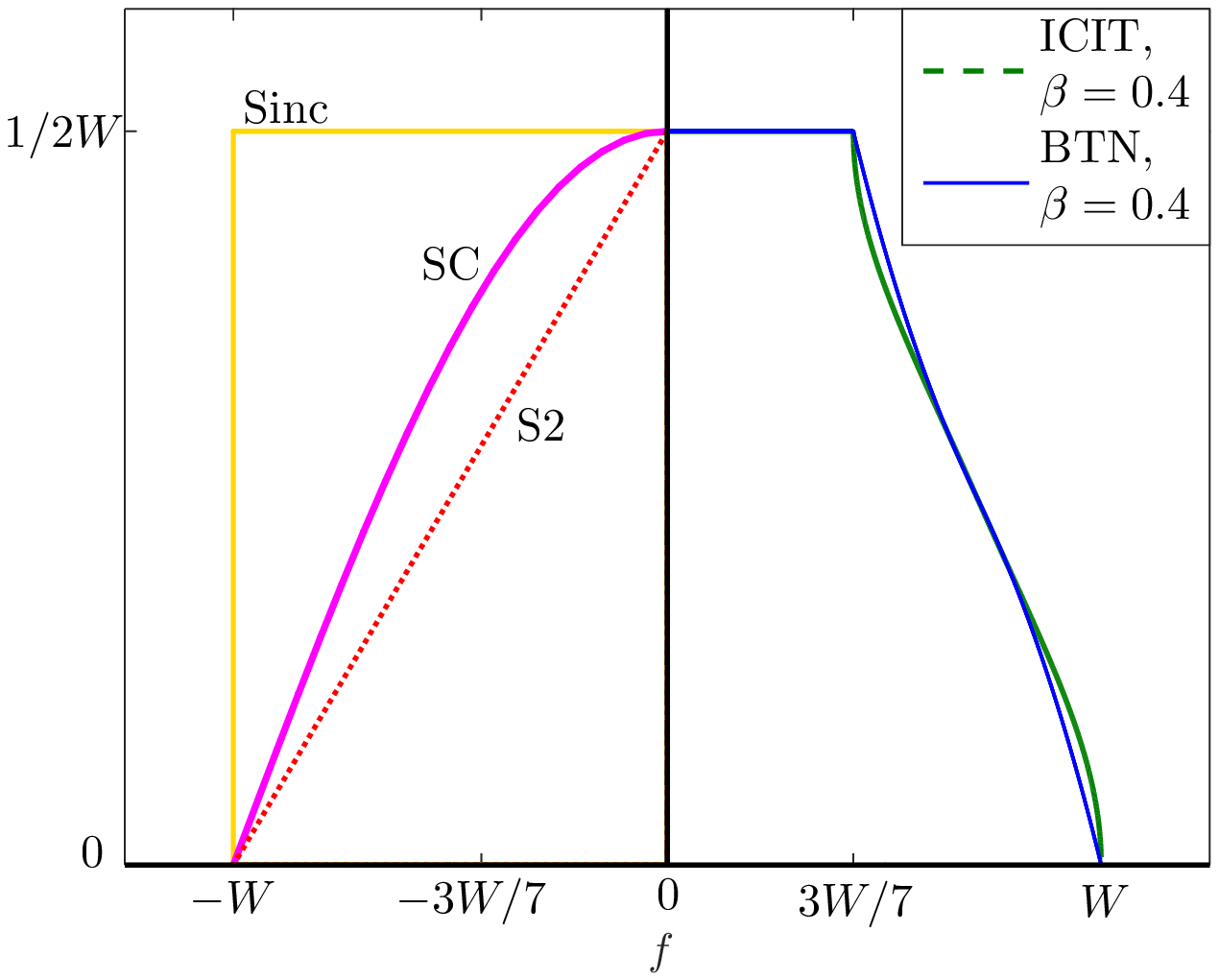}
\caption{Spectra of pulses considered in $\mspace{42mu}$ PAPR-BLOIC.}
\label{PulseOIC}
\end{minipage}
\begin{minipage}[t]{0.33\linewidth}
\centering
\includegraphics[width=2.48in,height=1.86in]{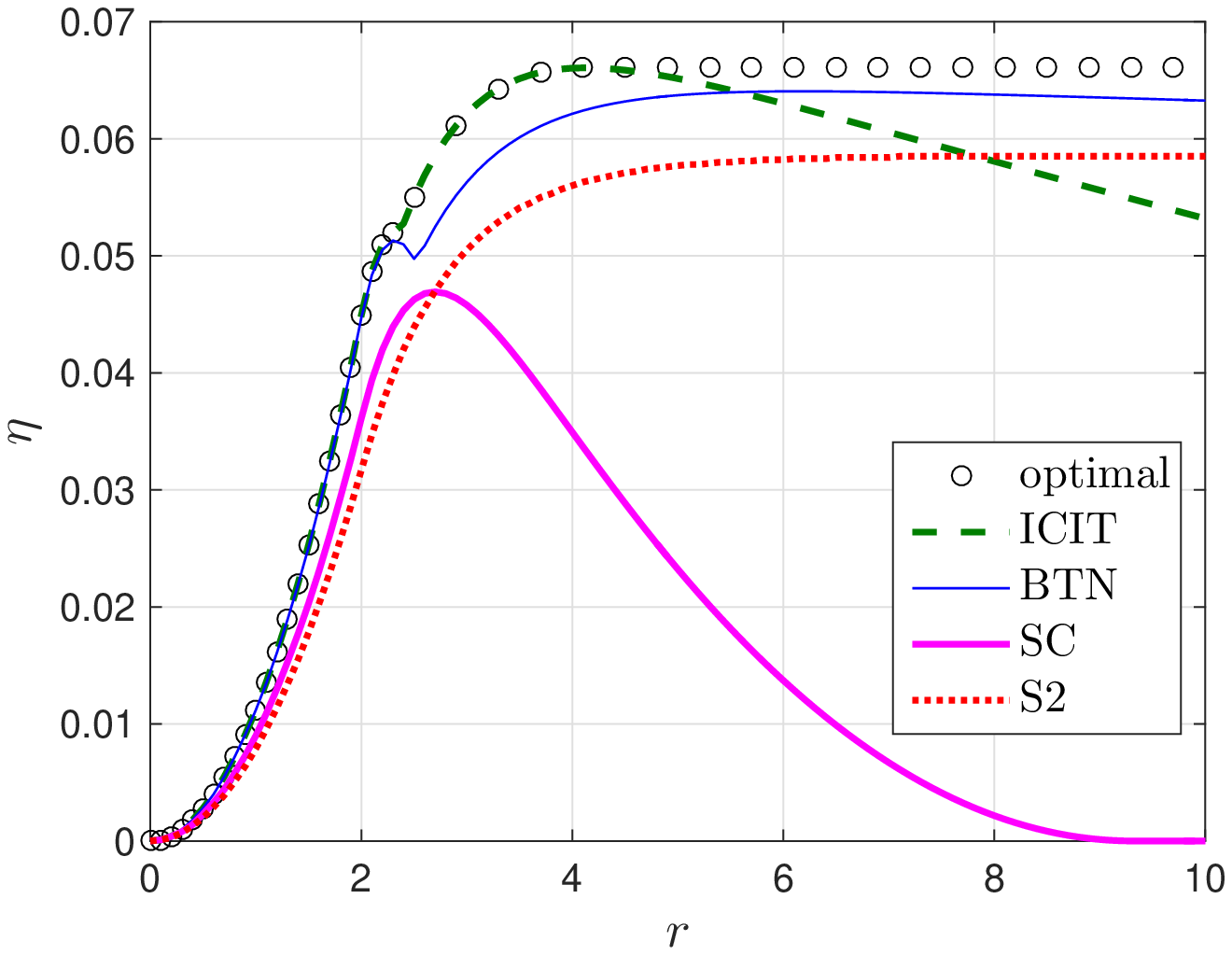}
\caption{Pre-OSNR factors of capacity lower $\mspace{35mu}$ bounds of PAPR-BLOIC.}
\label{OIC}
\end{minipage}
\begin{minipage}[t]{0.33\linewidth}
\centering
\includegraphics[width=2.48in,height=1.86in]{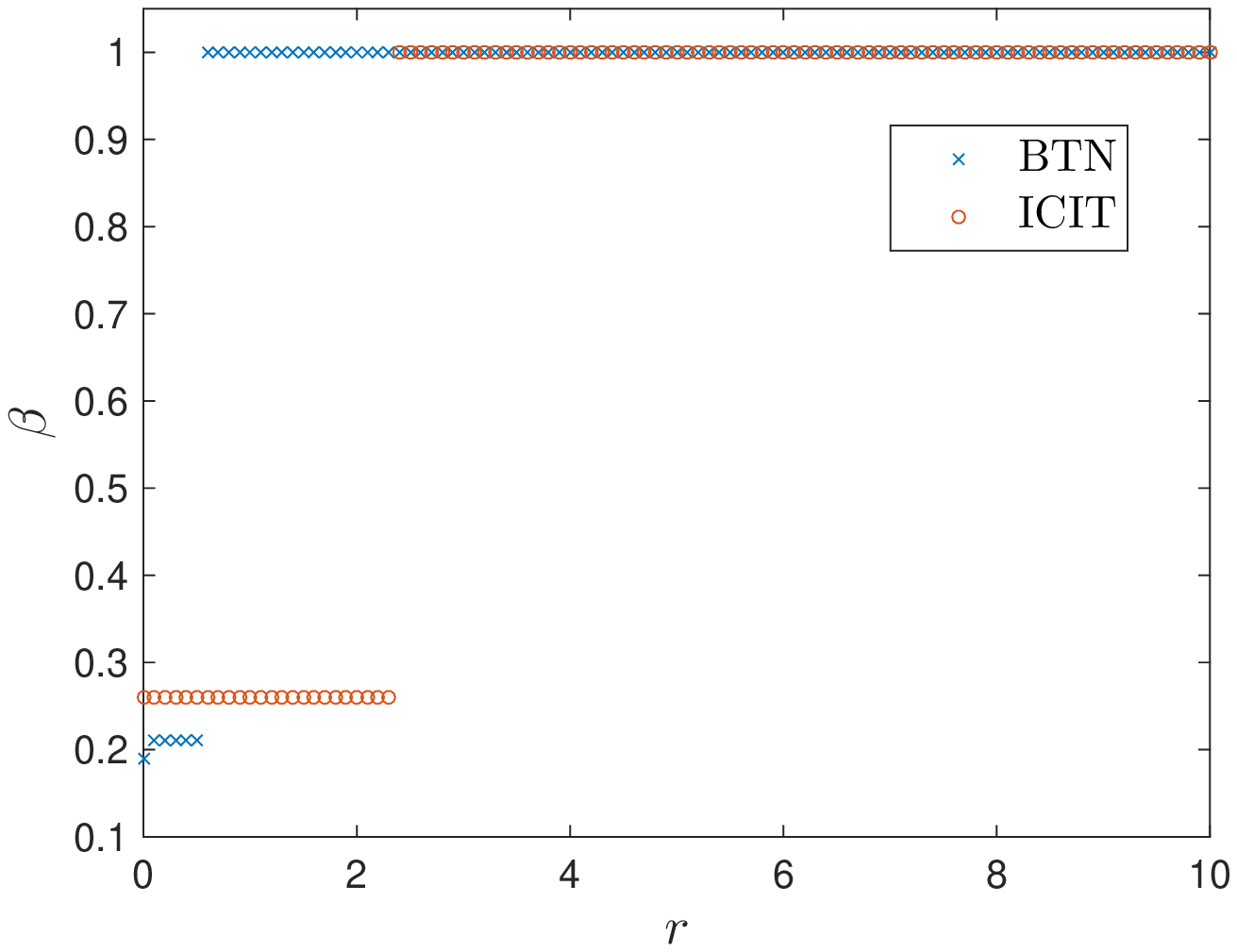}
\caption{Optimal value of $\beta$ for BTN and ICIT pulses in PAPR-BLOIC.}
\label{BetaOIC}
\end{minipage}
\end{figure*}
In intensity modulation and direct detection based optical communications,
a well-known channel model is the BLOIC,
\begin{equation}
\label{eqn:BLOIC}
Y(t)=S(t)+Z(t),\mspace{16mu}S(t)\ge 0,
\end{equation}
where $S(t)$ is the transmitted optical intensity (i.e., the optical power transferred per unit area), which is
bandlimited to ${\cal{W}}$ due to the limited available bandwidth of optoelectronic components, and $Z(t)$ is the same as that in (\ref{BLGC}).
We consider a PAPR constraint $r>0$ on the input optical power as
\begin{equation}
\label{E}
\lim_{T\rightarrow\infty}\frac{1}{2T} \mathbf E\left[\int_{-T}^T {S(t)\mathrm{d}t}\right]\leq {\cal E},\mspace{6mu} 0\leq S(t)\leq r{\cal E}.
\end{equation}

The capacity of the BLOIC has been studied in the literature; see \cite{HK04,HJC16,ZZ17} and references therein. In particular,
capacity bounds for the PAPR-BLOIC was derived in \cite{ZZ17} using the bounding techniques in Sec. II, and the lower bound is like
\begin{equation}
\label{LB-BLOIC}
C_{\textrm{PAPR-BLOIC}}(r)\ge W\log\left(1+\eta_g(r)\mathrm{OSNR}^2\right)\mspace{4mu}\textrm{bits/second},
\end{equation}
where $\mathrm{OSNR}\triangleq\frac {\mathcal {E}}{\sqrt{N_{0}W}}$, and the pre-OSNR factor is given by
\begin{equation}
\label{LB-BLOIC1}
\begin{split}
\eta_g(r)=
\begin{cases}
\frac { \mathcal G r^2}{2\pi e {\cal S}^2 }\left( \frac{e^{\mu}-1}{\mu e^{\mu}}\right)^2\exp\left(\frac{2 {\cal S}-r {\cal S}+r}{r}\mu\right),\mspace{4mu}r>2\\
\frac{\mathcal G r^2}{2\pi e {\cal S}^2},\mspace{172mu}0<r\leq2,
\end{cases}
\end{split}
\end{equation}
where ${\cal S}$ and $\mathcal G$ are defined as (\ref{S}) and (\ref{G}), respectively, with respect to an ${\cal L}_1$ pulse $g(t)$ normalized as
\begin{equation}
\label{nor}
\int _{-\infty} ^{\infty} g(t)\mathrm d t =\frac{1}{2W},
\end{equation}
and $\mu$ is the unique solution of
\begin{equation}
\frac{2{\cal S}-r{\cal S}+ r}{2 r}=\frac{1}{\mu}-\frac {1}{e^{\mu}-1}.
\end{equation}

In \cite{ZZ17}, although the benefit of pulse optimization had been recognized, only the S2 and SC pulses were used to evaluate the lower bound ({\ref{LB-BLOIC}}).
Now we further consider the BTN and the ICIT pulses in Table I.
Fig. \ref{PulseOIC} shows the spectra of these pulses which are normalized as (\ref{nor}) (note that in Fig. \ref{Pulse} the normalization is as (\ref{g})). Fig. \ref{OIC} gives the corresponding results on the pre-OSNR factor,
and Fig. \ref{BetaOIC} gives the corresponding optimal parameters. Compared to the results in Fig \ref{1}, the pre-OSNR factor lower bounds in Fig. \ref{OIC} behave differently.
The ICIT based lower bound performs the best at low PAPR (although it performs almost the same as the BTN based one when $r<2$), but it begins decreasing at $r\approx4.1$. The BTN
based lower bound decreases slower at high PAPR region, but our further calculation reveals that in fact it tends to zero as $r\to\infty$.
The S2 based lower bound is the only non-decreasing bound which tends to $\frac{1}{2\pi e}$.
Since a PAM waveform ensemble with a given PAPR is still admissible in channels with higher PAPR constraints, the pre-OSNR factor of capacity, $\eta(\mathrm{OSNR},r)$,
is always non-decreasing with $r$. So our optimal lower bound $\eta_\textrm{opt}(r)$ in Fig. \ref{OIC} is also non-decreasing:
\begin{equation}
\eta(\mathrm{OSNR},r)\ge \eta_\textrm{opt}(r)= \max\limits_{g}\max\limits_{r'\leq r}\eta_g(r').
\end{equation}
It was conjectured in \cite{ZZ17} that at high OSNR, the maximum achievable pre-OSNR factor using i.i.d. PAM ensemble in the AP-BLOIC is $\frac{1}{2\pi e}$,
which is achieved using the S2 pulse.
The present results show that the ICIT and BTN pulses perform better (but the input alphabet for $X$ must be carefully chosen),
and the former one achieves a pre-OSNR factor 0.06606, which outperforms $\frac{1}{2\pi e}$ by 0.26 dB. Moreover, this result also outperforms the result obtained by
a sphere packing argument in \cite{HJC16}, which shows that $\eta(\mathrm{OSNR})\ge 1/16=0.0625$ for the AP-BLOIC (i.e., PAPR-BLOIC as $r\to\infty$).

\section{Conclusion}
This paper studies the capacity of continuous-time Gaussian channels with given bandwidth under a PAPR constraint, namely, both a PP constraint and an AP constraint, connected by their ratio $r$.
Such a combination of constraints is more relevant to practical systems than a single constraint.
For both BLGC and BLOIC, by numerically evaluating information rates of i.i.d. PAM ensembles with optimized pulses, we obtain new achievable rate results which outperform existing ones. However, for both types of channels, the optimal pulse for a given PAPR is still unknown.
When $r$ is relatively small, the gaps between capacity upper and lower bounds are still considerable. These problems are left to future study.

\section*{Acknowledgment}
This work was supported in part by the Key Research Program of Frontier Sciences of CAS under Grant QYZDY-SSW-JSC003, by the National Natural Science Foundation of China under Grant 61722114, and by the Fundamental Research Funds for the Central Universities under Grants WK3500000003 and WK3500000005.

\end{document}